\documentstyle[prl,twocolumn,aps]{revtex}

\begin{document}

\draft

\title{
Surface tension
and dynamics of fingering patterns}

\author{
F.X. Magdaleno, J. Casademunt}
\address{
Departament d'Estructura i Constituents de la Mat\`eria\\
Universitat de Barcelona,
Av. Diagonal, 647, E-08028-Barcelona, Spain
}

\maketitle

\begin{abstract}

We study the minimal class of exact 
solutions of the Saffman-Taylor problem with zero surface tension,
which contains 
the physical fixed points of the 
regularized (non-zero surface tension) problem.
New fixed points are found and the basin of attraction of the Saffman-Taylor
finger is determined within that class.    
Specific features of the physics of 
finger competition are identified and quantitatively defined, 
which are absent in the zero surface tension case. This has  
dramatic consequences for the long-time asymptotics, revealing a   
fundamental role of surface tension in the 
\it dynamics \rm of the problem. 
A multifinger extension of microscopic
solvability theory is proposed to elucidate the interplay between 
finger widths, screening and surface tension.

\end{abstract}


\narrowtext

The displacement of a viscous fluid by a non-viscous one within the gap 
of a Hele-Shaw cell\cite{Bensimon}, 
has been one of the most studied problems in interfacial 
pattern formation for several decades\cite{reviews}. 
The relative simplicity of the problem has made possible an analytical
understanding of the subtle role of surface tension $\sigma$ in the selection 
of the Saffman-Taylor finger \cite{selection}, as a prototype
of the so-called microscopic solvability (MS) scenario of pattern 
selection\cite{reviews}. 
More recently, the increasingly interesting and controversial issue 
of the role of surface tension in the \it dynamics
\rm of fingering patterns has been raised.

It is known that the zero surface tension Saffman-Taylor (ST) problem is 
ill-posed as an initial value problem and is plagued of finite-time 
singularities\cite{Howison},\cite{Tanveer1}. Studies of laplacian growth
with zero surface tension, however, 
have proven insightful for instance in cases with needle-like 
growth\cite{needle}. 
In the case of smooth interfaces which concerns us here, 
a rich variety of physically relevant morphologies has been found 
among solutions of the $\sigma=0$ problem which remain smooth   
all the time (free of finite-time singularities)
\cite{Howison},\cite{Mineev1},\cite{Tanveer2}. 
Given the difficulty to get analytical information from the $\sigma \ne 0$ 
problem,  
this has raised the question of what part of the physics of 
fingering dynamics, if any, is captured by those solutions.
Within this spirit, Dai, Kadanoff and Zhou explored via numerical 
simulation the qualitative differences of distinct classes of initial
conditions\cite{Kadanoff}. 
For the so-called pole-like class and for finite-time, 
the problem was concluded to 
be qualitatively similar with and without 
surface tension.
More recently, Siegel and Tanveer\cite{Tanveer3} have shown 
that the regularized problem (i.e. vanishingly small $\sigma$) 
may differ significantly from the 
idealized problem ($\sigma=0$) in order one time, and therefore,  
smooth time evolving solutions of the $\sigma=0$ problem do \it not \rm
coincide, in general,  
with the limit of solutions of the regularized problem.
Evidence for this is shown in the particular 
class of single-finger configurations. This result, however, 
does not preclude other 
situations where this might no be the case\cite{footnote1}. 
More generally one can find situations 
where the evolution 
in the two cases is qualitatively equivalent in the sense 
that a (small) quantitative difference between the two remains
bounded for all time\cite{footnote5}. 
Further physical insight 
is thus still necessary to clarify the phenomenology which may be 
appropriately captured by the idealized problem, particularly 
concerning the long-time asymptotics.

From a different perpective, the MS scenario  
itself has been questioned recently by 
results of Ref.\cite{Mineev2} where it is claimed
 that in a wide class of nonsingular
exact solutions, the $\sigma=0$ dynamics leads naturally to the 
solution predicted 
by selection theory, without invoking surface tension to
explain selection. 
This seems to support the claim that surface tension is indeed
\it unessential \rm  
to the dynamics. In this letter we sustain the opposite 
conclusion\cite{footnote4}.

Our approach here consists of identifying 
specific dynamical features which can be viewed as essential to the
process of finger competition from a physical standpoint,
and check them in exactly solvable zero surface tension cases. 
This will allow for a precise diagnosis on the physical content
 of the idealized $\sigma=0$ 
problem. 
  
The evolution equation for the time-dependent conformal mapping $f(w,t)$ 
of the interior of the unit circle in the complex plane $w$ into the 
region occupied by the viscous fluid in the physical plane $z=x+iy$, 
 in the case of zero 
surface tension and in the co-moving frame (the frame  
moving with the mean interface 
velocity) can be written as 
\begin{equation}
\label{eq:eqmotion}
Re(w\partial_{w}{f(w,t)}[1+\partial_{t}{f^*(w,t)}])=-1.
\end{equation}
A channel of width $2\pi$ is considered along the $x$ axis, and 
periodic boundary conditions are assumed in the $y$ axis 
for simplicity. 

The basic idea is to find a simple, low-dimensional non-singular class 
of solutions of Eq.(\ref{eq:eqmotion}) which contains the physical 
fixed points of the regularized problem, and compare the 
phase space flow topology in both cases. 
The key point is that we do not need to know  
the exact
phase space trajectories of the regularized problem, which are 
particularly difficult 
to obtain even numerically for long times, but only the
phase space flow topology. The latter can indeed be inferred unambiguously 
from existing empirical 
evidence both experimental and simulational.

A class of solutions which satisfies the above requirements is of the form 
\begin{eqnarray}
\label{eq:ansatz}
f(w,t)  =  
 -\ln w + d(t) \nonumber\\
+ (1-\lambda) \left( \ln(1-\alpha(t)w) + \ln(1+{\alpha}^*(t)w) \right)
\end{eqnarray}
where $\alpha(t)=\alpha'(t)+i\alpha''(t)$.
This corresponds generically to two \it unequal \rm fingers.   
The interface 
shape has two symmetry axes along the fingers, separated  
a distance $\pi$. 
Such symmetry simplifies 
the analysis but does not affect the competition of fingers in
any fundamental way\cite{Casademunt}.

The case $\alpha'(t)=0$ corresponds to a single finger with 
the asymptotic ST shape for $\alpha'' \rightarrow 1 $.
The case  
$\alpha''(t)=0$ 
corresponds to two identical fingers which tend to a doubly degenerate 
ST solution as $\alpha' \rightarrow 1 $. 
For $|\alpha(t)| \ll 1$ the ansatz (\ref{eq:ansatz}) 
describes sinusoidal perturbations 
of a planar interface\cite{footnote2}.
$\lambda$ is a constant of motion and takes real values in the interval 
$[0,1]$.  
For well developed fingers $\lambda$ is the total filling fraction 
of the channel occupied by
the invading fingers.

For $\sigma=0$ this ansatz is exactly solvable in the sense that if we insert
Eq.(\ref{eq:ansatz}) into Eq.(\ref{eq:eqmotion})
we obtain a closed set of ordinary differential equations for the 
parameters $\alpha'$, $\alpha''$ and $d$. 
According to Ref.\cite{Mineev1} 
this case is free of finite-time singularities. 
The parameter $d$ accounts for a global displacement and is 
irrelevant for the present discussion.

For the sake of discussion and visualization, 
we find convenient to parametrize the phase space in terms
of the variables $u = 1 - \alpha''^2$ and $ r = 
(\alpha'^2+\alpha''^2-1)/(\alpha''^2 - 1)$.
Thus the phase space is the cube $[0,1]\times[0,1]\times[0,1]$ 
in the $(u,r,\lambda)$ space.
In these variables, the time evolution is given by the equations
\begin{eqnarray}
\label{eq:ode1}
\mathaccent95{u} = 2ru(1-u)  
\frac{3r - 4 - gr(1-ru)}{1 + gT_{g}(u,r)}\\
\label{eq:ode2}
\mathaccent95{r} = 2r(1-r)  
\frac{3r-2(1+ru)+g(1-ru)(2-r)}{1 + gT_{g}(u,r)} 
\end{eqnarray}
where
\begin{eqnarray}
\label{eq:ode3}
T_g(u,r) = (1-g)(2r+g(2r-1)) - \frac{1}{2}(1-g)^2ru \nonumber\\
-gur^2(1+g(ru-3))
\end{eqnarray}
and where $g=1-2\lambda=const$.

In order to compare with the physical case of $\sigma \ne 0$ we 
introduce the following construction. Consider a one dimensional 
set of initial conditions ($t=0$) of the form Eq.(\ref{eq:ansatz}) 
surrounding the planar interface (PI) fixed point $u=1$, $r=1$,  
for a fixed $\lambda$. We take them 
infinitessimally close to PI in such a way that the interface is in the 
linear regime\cite{footnote2}. 
The time evolution from $t=-\infty$ to $t=\infty$ of this set 
spans a compact two dimensional phase space $(u',r')$ imbeded in the 
infinite dimensional space of interface configurations. 
For a finite surface tension,  
we know from all existing empirical 
evidence that the above subspace must contain three fixed points,
namely the (unstable) planar interface (PI), 
the (stable) ST single finger (1ST)
and a saddle fixed point corresponding to the degenerate double ST finger
(2ST). 
For finite $\sigma$ (and in the high viscosity contrast limit
\cite{Casademunt}),
 the 1ST fixed point is known to be the universal attractor of this problem 
so all trajectories start at PI and end up at 1ST. 
The 2ST fixed point has a lower dimension attracting manifold defined 
by $\alpha''=0$ and   
 will govern the dynamics of finger competition.
We may define the space $(u',r')_0$ as the limiting case of 
$\sigma \rightarrow 0$ (taken after the limits $t \rightarrow \pm \infty$).

Since in the linear regime the regularized problem for vanishingly 
small $\sigma$ converges regularly to the $\sigma=0$ solution, 
the manifolds $(u,r)$ and $(u',r')$ must be tangent at the PI fixed point, 
$u=1, r=1$ (see Fig.1).
Furthermore, $(u,r)$ and $(u',r')_0$ 
must intersect at 1ST and 2ST  
which occur respectively at $u=0, r=1$ and  
$u=1, r=0$, as seen directly from
Eqs.(\ref{eq:ode1})-(\ref{eq:ode3}).
Although $(0,1)$ and $(1,0)$ are fixed points for any
$\lambda$, according to selection theory the intersection will occur 
at $\lambda=1/2$.

Following the topological approach of Ref.\cite{Casademunt}, 
it is useful to consider the stream function $\psi$, 
defined as the imaginary part of the complex potential
$\Phi(w,t) = -f(w,t) -\ln w $, in the co-moving frame. 
Along the interface, $\psi$ is then a periodic function which provides 
a natural definition of individual aerial growth rate of fingers, 
which we call $\Delta \psi_L$ and $\Delta \psi_S$ 
for the longer and shorter fingers respectively. 
In this simple case these are given 
as maximum-to-minimum differences of 
the stream function extrema along the interface.
$\psi$ may have only one maximum even for two-finger configurations, 
in which case we take $\Delta \psi_S=0$ and qualify the finger 
as 'non-growing'.
In our case, and for finite width fingers, 
$\Delta \psi > 0$ (growing) and $\Delta \psi = 0$ 
(non-growing) correspond respectively to positive and negative 
tip velocities relative to the mean interface position.

The physical scenario of finger competition which we want to test,
extracted from experiments 
and simulations\cite{Casademunt} 
can be briefly described as follows.
In the linear and early nonlinear 
regimes two different fingers grow 
with both $\Delta \psi_L(t)$ and $\Delta \psi_S(t)$ increasing with time. 
When the fingers are well developed and the 'growth' function $G(t)=
\frac{\Delta \psi_L(t) + \Delta \psi_S(t)}{1-\lambda}$ (with $1-\lambda=
\Delta \psi_L(\infty)+\Delta \psi_S(\infty)$)
is of order one, 
the competition takes over. This is signaled by an enhanced growth of the 
'competition' function $C(t)=\frac{\Delta \psi_L(t)-\Delta \psi_S(t)}
{\Delta \psi_L(t)+\Delta \psi_S(t)}$ as 
$\Delta \psi_L$ starts to increase 
at the expense of $\Delta \psi_S$. Existence of competition can thus be 
identified with $\Delta \psi_S$ decreasing with time. The competition may 
be termed 'succesful' when $\Delta \psi_S \rightarrow 0$ asymptotically 
($C(\infty)=1$), that is when a 'growing' finger is turned into 'non-growing' 
due to the presence of another finger.
As discussed in Refs.\cite{Casademunt},
 this  dynamical elimination of the small finger  
is associated to topology changes in the physical 
velocity field, which occur via the crossing of topological defects 
through the interface. 

We now discuss the zero surface tension dynamics of our ansatz 
(\ref{eq:ansatz}), 
by analyzing the phase portrait of the dynamical system defined 
by Eqs.(\ref{eq:ode1})-(\ref{eq:ode3}). This is plotted in Fig.1  
for $\lambda=1/2$. 
In this case Eqs.(\ref{eq:ode1})-(\ref{eq:ode3})
can be integrated analytically. Dynamical trajectories are of the form 
\begin{equation}
\frac{2u - 3ru + r^2u^2}{\sqrt{u(1-r)(1-ru)}}=const
\end{equation}
and are plotted as solid lines with arrows. The dashed lines are 
$\lambda$-independent boundaries. The short-dashed line separates 
the one-finger (above) and the two-finger (below) regions. The long-dashed 
one is the defect boundary separating 
the no-defect (above) and the 1-defect (below) regions, 
with $\Delta \psi_S=0$ and $\Delta \psi_S > 0$ respectively.

Our central result is that the topological structure 
of the phase portrait for $\sigma=0$ (Fig.1) 
is radically different from that of finite surface tension.
Its most salient feature is the fact that the ST single-finger 
solution has a limited basin of attraction. Part of the flow 
evolves towards a continuum of (attractive) fixed points $r=0$.
The separatrix of the two regions is a critical trajectory
ending in a new (saddle) fixed point located at $u^{*}=0$
and $r^{*}=2\lambda/(1+\lambda)$.

For arbitrary $\lambda$, the line $r=0$ 
is a continuum of 
stationary solutions with coexisting \it unequal \rm fingers (different widths 
$\lambda_1$, $\lambda_2$ with $\lambda_1+\lambda_2=\lambda$) 
advancing with the \it same \rm velocity and with
tip positions separated in the $x$-direction 
by a finite distance $\Delta=(1-\lambda)\log\frac{1+\sqrt{1-u}}{1-\sqrt{1-u}}$.
Solutions of this type have been 
reported previously\cite{Mineev1},\cite{Tanveer2}. 
We would like to call the attention upon the fact that 
the screening of the laplacian field, 
as the mechanism usually invoked to explain competition, applies actually to 
the aerial growth of the small finger $\Delta \psi_S$, which is reduced 
indeed by the presence of the longer neighboring finger, 
but not to the velocity, 
which may be in fact the same. Concepts such as screening length and time 
\cite{needle} are 
therefore meaningful only if the widths of competing fingers are 
constrained to be equal. 

The fixed point $(u^*,r^*)$ corresponds to a 
new type of asymptotic stationary solution of the $\sigma=0$ ST problem.
It consists of  
two fingers with unequal positive velocities. The length ratio of the 
fingers satisfies\cite{Magdaleno1}
 $\lim_{t\rightarrow\infty}L_S/L_L=1/3$ independently of
$\lambda$. 
For the $r=0$ solutions we have $\lim_{t\rightarrow\infty}L_S/L_L=1$ 
while, for the 1ST fixed point when approached from the two-finger region 
we have $\lim_{t\rightarrow\infty}L_S/L_L=0$. In the latter case, 
the residual non-growing finger which subsists is reminiscent of 
the 'frozen' fingers observed in real experiments.
 
Our central point is that, according to the above discussion, 
the possibility of successful competition is associated to  
the fact that dynamical trajectories cross the defect boundary from below 
(annihilation of topological defects\cite{Casademunt}).
In Fig.1 we see that,  
for $\lambda=1/2$, there is no successful competition whatsoever 
since the critical 
trajectory is located above the defect boundary\cite{footnote3}. 

The cases of $\lambda \neq 1/2$ and $\sigma=0$ are not directly relevant to 
the viscous fingering problem, but may
be relevant to other generic situations of laplacian growth in the 
spirit of Refs.\cite{needle}, 
and will be discussed in detail 
elsewhere\cite{Magdaleno1}. Here we will just remark that  
$r^{*}(\lambda)$ is monotonically increasing, between  
$r^{*}(0)=0$ and $r^{*}(1)=1$. 
Therefore, the basin of attraction of the single finger solution 
is larger for 
narrower fingers. Furthermore, since the defect boundary is independent 
of $\lambda$, there exists a critical $\lambda_c=1/3$ for which  
$r^{*}$ crosses the defect boundary. This implies that, for 
$\lambda < 1/3$, there are dynamical trajectories which cross 
the defect boundary from below, and therefore the competition is then 
successful 
for some finite region of phase space. 

In summary, from the analysis of the $\sigma=0$ dynamics of the class 
Eq.(\ref{eq:ansatz}) 
we conclude that (i) only a small $\lambda$-dependent 
part of phase 
space behaves qualitatively as the $\sigma \neq 0$ problem, leading 
to a ST single finger (with maybe a residual non-growing finger); 
(ii) dynamical elimination of growing fingers does not occur for  
finger widths relevant to the problem of viscous fingering ($\lambda=1/2$);
(iii) the picture of competition based solely on laplacian screening 
is insufficient, since relative widths of fingers and not only 
relative tip positions come into play. 

In order for the 'screening' picture to be valid, 
an additional dynamical constraint
is required to force the finger widths to be equal.
In growth processes based on aggregation of particles, the 
finger width may be fixed by particle size (set to zero in Refs.\cite{needle}).
In the problem of viscous fingering such constraint is supplied precisely
by surface tension.
This suggests that an extension of MS, which is essentially a
\it static \rm theory, to multifinger configurations,  may shed new light 
on the \it dynamics \rm of the problem.

The generalized multifinger MS scenario can be sketched as follows. For  
two-finger configurations, there exists a two parameter
continuum family of steady state solutions which we can parametrize 
by $\lambda=\lambda_1+\lambda_2$ and $p=\lambda_1/(\lambda_1+\lambda_2)$.
The cases $p=0,1$ correspond trivially to the single-finger case. The case 
$p=1/2$ (two identical fingers) is also reducible to single-finger MS
in a channel of half width. Most interestingly,  
one can show\cite{Magdaleno2} that, for nonzero surface tension,
 nontrivial stationary 
solutions with \it unequal \rm fingers ($p \ne 1/2$) exist. 
In this case, surface tension
selects an infinite set of values of 
$\lambda$ which differs from the single-finger case, but which scale also 
as $(\lambda-1/2) \sim \sigma^{2/3}$. 
Furthermore, 
for any given $\lambda$ of the above discrete set, 
there exists another countably infinite 
set of possible values of $p$ with 
$(p-1/2) \sim \pm \sigma^{1/3}$\cite{Magdaleno2}.
This new set of fixed points of the problem with surface tension 
and its physical relevance will be discussed in detail 
elsewhere\cite{Magdaleno2}. 
It is reasonable to expect that all two-finger solutions except the 
$p=1/2$, $\lambda=\lambda_m(\sigma)$ (with $\lambda_m$ the minimum value 
of $\lambda$ from MS) 
lie outside the space $(u',r')$.  
Despite the fact of being globally unstable, the 
$p=1/2$, $\lambda=\lambda_m$ fixed point is then the physically relevant 
one to describe finger competition since it has an attracting manifold 
which includes the PI fixed point (linear regime). 
The process of finger competition 
can thus be pictured as follows. From the linear instability 
a given number of fingers emerge. As far as this early stage is dominated by 
the most linearly unstable mode (in the limit of weak white noise on
PI, the emerging configuration will indeed be nearly periodic),
the interface is relatively close to the attracting manifold of the 
nST fixed point (n equal fingers with $\lambda_i=\lambda_m(\sigma)/n)$. 
The fingers tend thus to adopt the same fingertip curvature 
and select their widths at early stages of the nonlinear regime 
according to single-finger MS theory.
The nST fixed will then govern the process of competition in the 
sense that the path
connecting typical initial configurations  
with the single finger attractor, must necessarily 
pass near that saddle point.   
The phenomenon of competition is then viewed as the crossover to the unstable 
directions of the nST, $\lambda=\lambda_m$ fixed point. 
Such crossover is what is missed in the $\sigma=0$ problem, since the 
unstable direction of the 2ST fixed point becomes infinitelly marginal
(a line of fixed points) in that limit.
In the terminology of dynamical systems, this reflects 
the fact that Eqs.(\ref{eq:ode1})-(\ref{eq:ode3})
are structurally unstable\cite{Peixoto}.
We thus conclude that surface tension plays 
a fundamental role in the \it dynamics \rm of finger competition and that,
for the long time asymptotics,  
it can only be treated as a 'regular' perturbation in a very limited 
region of phase space which excludes multifinger configurations. 

We are indebted to David Jasnow for stimulating discussions.
We acknowledge financial support from the Direcci\'on General de
Ense\~{n}anza Superior (Spain), Project PB96-1001-C02-02, and 
the European Commission Project ERB FMRX-CT96-0085.
F.X.M. also acknowledges financial support from the Comissionat
per a Universitats i Recerca (Generalitat de Catalunya).

Fig 1. Phase space flow of the dynamical system defined by 
Eqs.(\ref{eq:ode1})-(\ref{eq:ode3}) for 
$\lambda=1/2$. See explanation in text.


\begin{references}
\bibitem{footnote1}
For instance, a sequence of solutions of $\sigma \ne 0$ can be found which 
converges uniformly for all time to the $\lambda=1/2$ time-dependent 
solution with $\sigma = 0$.
\bibitem{footnote4}
Our results are not really incompatible with a proper interpretation of
the results of Ref.\cite{Mineev2} and their implications\cite{Magdaleno1}.
\bibitem{footnote3}
Strictly speaking there may be a certain amount of competition in the
sense of a decrease of $\Delta \psi_S(t)$ but quantitatively very small.
\end{references}
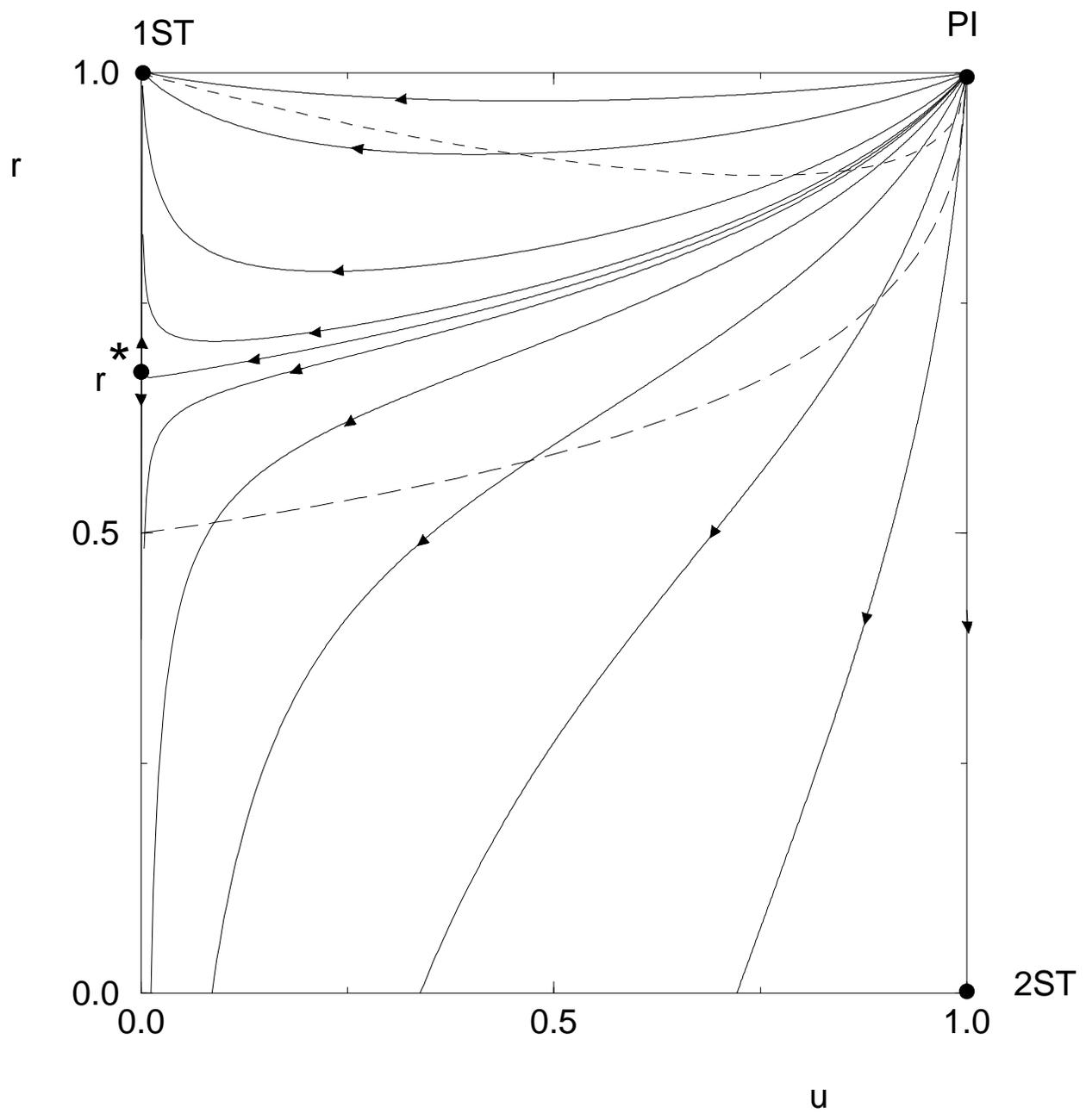
\end{document}